\documentclass[aps,prb,twocolumn,superscriptaddress,groupedaddress]{revtex4}
\usepackage{amsmath,amssymb,color,graphicx,hyperref,amsthm,bm,epstopdf,mathrsfs}
\usepackage{hyperref}
\usepackage{graphicx,grffile}
\usepackage{amsmath}
\usepackage{amsfonts}
\usepackage{comment}
\usepackage{array}
\usepackage{url}

\usepackage[toc,page]{appendix}

\renewcommand{\ol}{\overline}

\newcommand{\area}{\mathcal{A}}

\renewcommand{\eqref}[1]{eq. \ref{#1}}
\newcommand{\figref}[1]{fig. \ref{#1}}

\definecolor{TW-colour}{RGB}{100,0,100}
\definecolor{Error-colour}{RGB}{250,0,0}
\definecolor{Referee-colour}{RGB}{225,25,25}

\makeatletter
\let\cat@comma@active\@empty
\makeatother

\begin{document}

\title{Magnetic response of nematic superconductors: skyrmion stripes and their signatures in muon spin relaxation experiments}
\author{Martin Speight}
\affiliation{School of Mathematics, University of Leeds, Leeds LS2 9JT, United Kingdom}
\author{Thomas Winyard}
\affiliation{Maxwell Institute of Mathematical Sciences and School of Mathematics, University of Edinburgh, Edinburgh, EH9 3FD, United Kingdom}
\author{Egor Babaev}
\affiliation{Department of Physics, KTH-Royal Institute of Technology, Stockholm, SE-10691 Sweden}

\begin{abstract}

We investigate the magnetic response of nematic superconductors, presenting a new approach to find vortex and skyrmion structures beyond symmetry-constraining ans\"atze. Using this approach we show that nematic superconductors form distinctive skyrmion stripes. Our approach lends itself to accurate determination of the field distribution for muon spin rotation probes. We use this to show that the skyrmion structure manifests as a double peak in the field distribution, markedly different from the signal of standard vortex lattices.
\end{abstract}

\maketitle

Understanding nematicity in superconducting states is currently one of the central questions of the field \cite{fu2010odd,fu2014odd,venderbos2016odd,wang2021progress,zyuzin2017nematic,wu2017majorana,chirolli2018chiral,barkman2019antichiral,
uematsu2019chiral,dentelski2020effect,yiphalf,
chubukovTBG,cho2020z3}. 
Various indications of nematicity have been observed in a broad range of materials including cuprates \cite{hinkov2008electronic,daou2010broken}, heavy fermion materials \cite{ronning2017electronic,okazaki2011rotational} and iron based superconductors \cite{fernandes2022iron,wang2023nematicity,chu2012divergent, lu2014nematic,kuo2016ubiquitous}. One particular example of a material that has been the target of intense experimental study is the doped topological insulator $\mathrm{Bi_2 Se_3}$
\cite{hor2010superconductivity,wray2010observation,kriener2011bulk}. Superconducting candidate materials of interest include $\mathrm{Cu_x Bi_2 Se_3}$, $\mathrm{Sr_xBi_2Se_3}$ 
\cite{Shruti_SrBiSe,Liu_SrBiSe,Pan_Hc2,Matano.Kriener.ea:16,tao2018direct} and $\mathrm{Nb_xBi_2Se_3}$ \cite{Asaba_NbBiSe}.
One of the interesting properties of nematic superconductors is that the flux carrying objects could be skyrmions \cite{nematic,zyuzin2017nematic,how2020half}, which are fundamentally distinct from the vortices that normally appear in superconductivity.

The appearance of skyrmions raises two questions: how do the skyrmions affect the magnetic response of the material, and what unique experimental signatures does this produce? The first question is technically nontrivial, since the inter-skyrmion interactions are complicated and anisotropic, driven by mode mixing in general multicomponent anistropic models \cite{silaev2018non,winyard2019hierarchies,winyard2019skyrmion,speight2019chiral,nematic}, which dictates the appearance of multiple scales in the inter-skyrmion forces.  This anisotropic behaviour cannot be captured by a simple ansatz, thus it is challenging to approximate the resulting lattice with any certainty. Using brute-force numerics, on the other hand, would require simulating complicated Ginzburg-Landau models at large system sizes, with boundaries that may affect structure formation.

In this work and in the companion extended paper \cite{nematic} we study this magnetic response by developing and applying a new method that can determine vortex lattice structure for any model. We find that nematic superconductors exhibit an unusual magnetic response in the form of skyrmion stripes.

Since surface probes visualize vortex structures at limited scales and structure formation may be affected by pinning, strain,inhomogeneities and surface effects, the most reliable method to detect these structures is muon spin rotation ($\mu SR$) experiments. These fire muons into the sample, detecting the orientation of positrons emitted when the muons decay. This data can then be used to calculate the statistical distribution of the   magnetic field {\em in the bulk} of a superconductor \cite{sonier2000musr}.

We will demonstrate that for nematic superconductors, this magnetic field distribution exhibits a double peak structure. We posit that the number of peaks in the distribution correlates with the number of length scales in the lattice. For a nematic superconductor this is two, the separation between chains and the inter-skyrmion separation along the chain, which we will call the chain link length.

The model we consider in this paper is derived from the proposal that certain topological insulators, such as electron-doped 
nematic superconductors (e.g. $Cu_x Bi_2 Se_3$) are described by a two-component order parameter \cite{hor2010superconductivity,wray2010observation,kriener2011bulk,fu2010odd}. This order-parameter breaks rotational symmetry in the basal plane to a mixed symmetry, combining the phase difference of the order-parameter and spatial rotations \cite{nematic}. This matches the observed two fold rotational symmetry of specific heat and upper critical field \cite{venderbos2016identification}. 
The associated Ginzburg-Landau Gibbs free energy for such a nematic superconductor was microscopically derived in \cite{zyuzin2017nematic}. In dimensionless units, it reads
\begin{equation}
G = \int_{\Omega}\left\{\frac{1}{2} Q^{\alpha\beta}_{ij} \ol{D_i \psi_\alpha} D_j \psi_\beta + \frac{1}{2} |B-H|^2 + F_p(|\psi_\alpha|, \theta_{\alpha\beta})\right\},
\label{eq:G}
\end{equation}
where $\psi_\alpha = \rho_\alpha e^{i\varphi_\alpha}$ are  two complex superconducting order parameters $\alpha \in \{1,2\}$. Note that Greek indices will always be used for superconducting components, Latin indices will be used for spatial directions and repeated indices imply summation. $F_p$ denotes the collected potential terms and varies between models, but we will use the standard form,
\begin{equation}
F_p = \frac{\eta^2}{2}\left(-|\psi_1|^2 - |\psi_2|^2 + \frac{1}{2}(|\psi_1|^4 + |\psi_2|^4) + \gamma|\psi_1|^2 |\psi_2|^2\right),
\label{eq:Fp}
\end{equation}
where $\gamma$ a microscopically derived parameter \cite{zyuzin2017nematic} related to the nematic nature of the system. It directly determines the ground state value for the condensates and is set to $\gamma = 1/3$ according to \cite{zyuzin2017nematic}. The parameter $\eta$ scales the potential and hence the ratio of critical fields $H_{c_2}/H_{c_1}$. Note that for $\eta << 1$ the system starts to act like a type-I superconductor and skyrmions do not form.

The constant matrices $Q^{\alpha\beta}$ define the anisotropy in the system and must obey the symmetry $Q^{\alpha\beta}_{ij} = \overline{Q}^{\beta\alpha}_{ji}$, ensuring that the energy is real. $D_i = \partial_i - iA_i$ is the covariant derivative, where $A$ is the vector potential corresponding to a $U(1)$ gauge field, with corresponding magnetic field $B = \nabla \times A$. For a nematic superconductor the anisotropy matrices take the form
\begin{align}
Q^{11} = Q^{22} = \left( \begin{array}{ccc} 1 & 0 & 0 \\ 0 & 1 & 0 \\ 0 & 0 & \beta_z\end{array}\right), \quad Q^{12} = \beta_\perp\left( \begin{array}{ccc} 1 & i & 0 \\ i & -1 & 0 \\ 0 & 0 & 0 \end{array}\right),
\label{eq:nematic}
\end{align}
where $\beta_z$ and $\beta_\perp$ are positive, we will follow \cite{zyuzin2017nematic} with the choice $\beta_z = 4/3$ and $\beta_\perp = 1/3$.  The $\beta_\perp$ term drives skyrmion formation or fractional vortex splitting. This can seen by considering co-centred fractional vortices where the term integrates to zero. Hence by splitting the fractional vortices this term can become negative,  thus if $\beta_\perp$ is sufficiently large it is energetically favourable to have fractional vortex splitting.  The field equations in the bulk of the superconductor can then be found by varying the Gibbs free energy w.r.t.\ the fields $\psi_\alpha$ and $A_i$.  Note that the resulting equations are independent of $\beta_z$ which will have no effect on our results.

\begin{figure}
\includegraphics[width=0.4\linewidth]{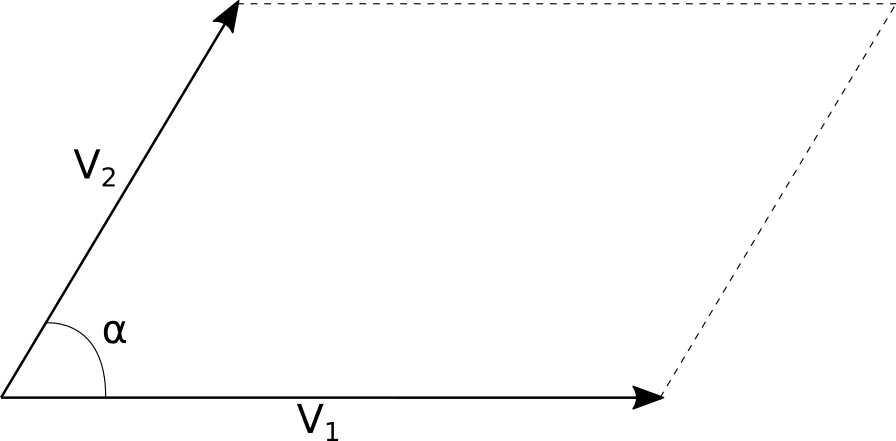}
\caption{\label{fig:unitCell}
Diagram of the geometry of a general unit cell for a vortex lattice, defined by the two vectors $V_1$ and $V_2$ with angle $\alpha$.}
\end{figure}

Solutions for isolated flux-carrying objects show that in nematic superconductors these are skyrmions, that have the form of spatially separated half-quantum vortices \cite{zyuzin2017nematic,how2020half,nematic}. These solutions were found by simulating the field equations on a plane ($\Omega = \mathbb{R}^2$), however to find a lattice solution, we must solve them over the unit cell of a lattice with periodic boundary conditions (up to gauge). While it is often assumed that the unit cell has hexagonal symmetry, there is no reason to expect this for skyrmions in an anisotropic model. Hence we take a new approach, considering a general unit cell, which takes the form of a parallelogram that is periodic (up to gauge), as shown in \figref{fig:unitCell}. A lattice solution is then given by the unit cell and field configuration that minimises $G/\mathcal{A}$, where $\mathcal{A}$ is the area of the unit cell. Hence numerically, we seek minimisers of $G/\mathcal{A}$ w.r.t. the fields and w.r.t. the geometry of the unit cell itself, represented as the two vectors $V_1$ and $V_2$ in \figref{fig:unitCell} (note that this will also control the area $\mathcal{A}$). The full technical detail of the procedure we propose, is given in a companion paper   \cite{nematic} (along with a generalisation that is not of use in this paper). A brief account is also presented in the appendix.

By tessellating the solution on the periodic unit cell, we find a solution that covers the plane with a vortex lattice. Hence, in applying the above method we find the field configuration that minimises $G/\mathcal{A}$ in the bulk of a superconductor with no boundary effects. In addition, as we know the shape of the unit cell, we know the symmetry of the lattice.

\begin{figure*}
\includegraphics[width=0.8\linewidth]{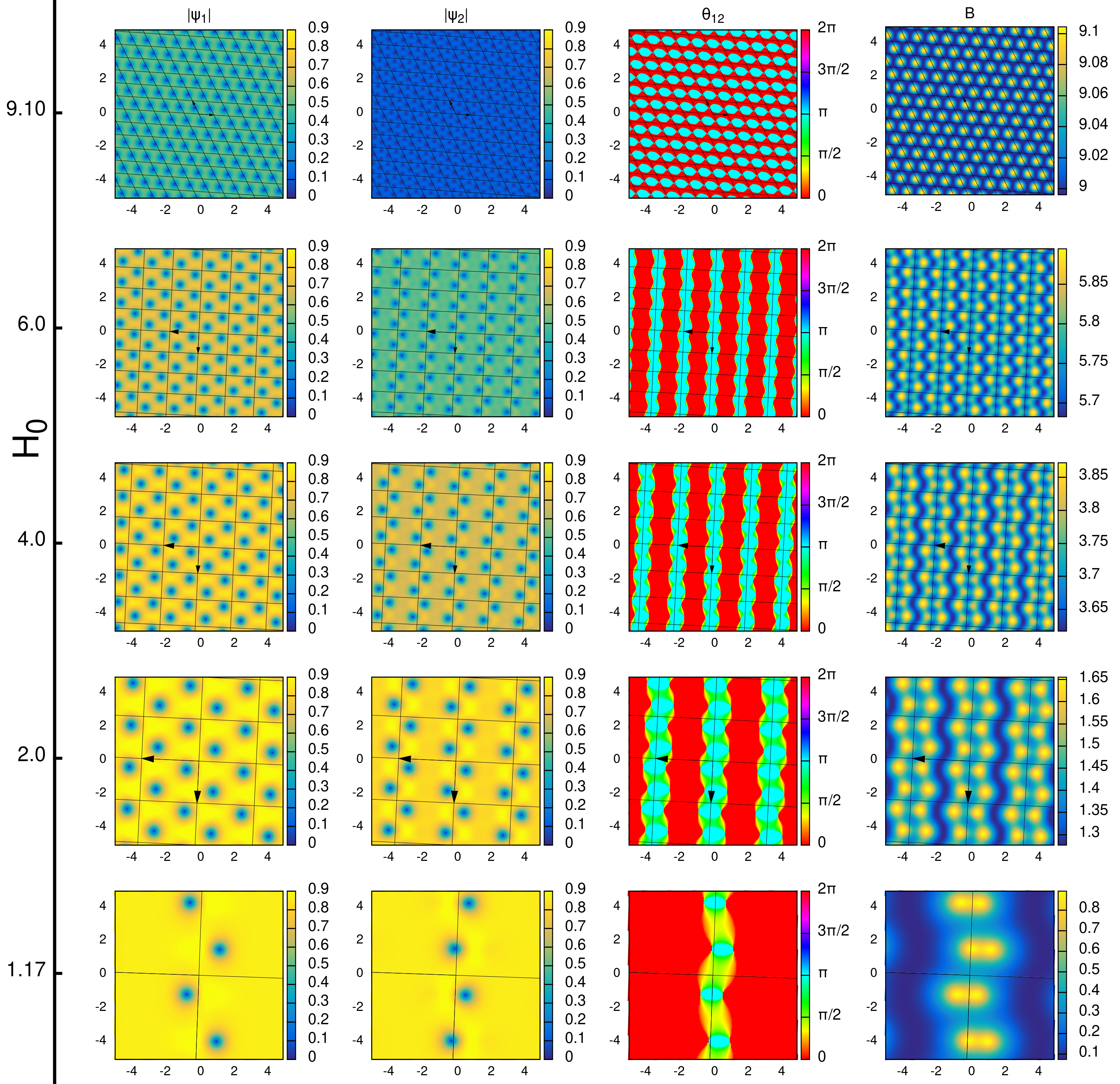}\caption{\label{fig:lattice}
Contour plots of the lattice solutions (skyrmion chains) for the model \eqref{eq:G} in the Basal plane, with $\eta = 3$ and increasing external field $H$. The fields plotted are the order parameter magnitudes $|\psi_1|$ and $|\psi_2|$, the phase difference between the two complex order parameters $\theta_{12} = \varphi_1 - \varphi_2$ and the local magnetic field strength $B$. The lattice consists of skyrmion chains, formed by rectangular unit cells (marked with black lines) containing four half-quantum vortices or winding two. As the external field increases the skyrmion chains squash together. Then near $H_{c_2}$ a transition occurs and the vortices are forced into the conventional triangular lattice with unit cells of winding one. Note that there is no simple symmetry between $|\psi_1|$ and $|\psi_2|$ as the symmetry is broken to a mixed symmetry with spatial rotations \cite{nematic}.}
\end{figure*}

Applying the above approach for the basal plane of a nematic system gives the results in \figref{fig:lattice}, where we see that for low and medium strength external field $H$, the unit cell is rectangular ($\alpha = \pi/2$), containing four half-quantum vortices (totalling winding two) in each cell, forming a Skyrmion chain. The lattice vectors lie parallel to the chain and perpendicular to the chain. Hence, we can understand such a nematic unit cell by considering two lengths:
\begin{itemize}
\item $|V_1|$ - chain link length,
\item $|V_2|$ - the distance between the chains.
\end{itemize}
We can see how these values change in relation to an increasing external field strength in \figref{fig:lattice}. As the strength of the external field $H$ is initially increased, the density of chains increases and $|V_2|$ decreases. Then we see that $|V_1|$ also decreases as the chain segments are squashed. Eventually when the external field $H$ becomes large and approaches $H_{c_2}$, the vortices are forced together and form the familiar tightly packed triangular lattice. Note that $V_1$ is related to the fractional vortex separation of the model which is driven by $\beta_\perp$, and both $V_1$ and $V_2$ will approximately scale inversely with $\eta$ (the exact relation is nonlinear).

Theoretically in clean samples these skyrmion structures should be visible on the surface using scanning SQUID or scanning Hall probes. However, flux structure near the surface can be affected by surface physics and it is difficult to probe large areas. Hence,
one of the most powerful tool to identify skyrmions is the muon spin relaxation technique, which gives statistical information regarding the flux distribution in the bulk of the system.
To compare the lattice configurations in \figref{fig:lattice} with $\mu$SR experiment, we must convert the numerical field configuration for a given unit cell to a probability density distribution $p(B)$. Our method is ideal for this as it produces the field configuration on a single unit cell, with no boundary effects and no deformations. We will use a standard kernel density estimation \cite{parzen1962estimation} to approximate a continuous distribution from our discrete field configuration. Namely, our numerical algorithm produces a discrete configuration,  where the field distribution is a set of normalised delta functions. We choose to smooth these delta functions by replacing them with localised Gaussians and taking the normalised sum. Hence, the probability density of field strength $B$ in the unit cell is,
\begin{equation}
p(B) = \frac{1}{n_1 n_2 h} \sum^{n_1}_{i=1}\sum_{j=1}^{n_2} \frac{1}{\sqrt{2\pi}} e^{-\frac{1}{2}  (\frac{B - B_{ij}}{h})^2},
\label{eq:probability}
\end{equation}
where $B_{ij}$ is the local magnetic field at lattice site $(i,j)$ with $i\in[1,n_1]$ and $j \in [1,n_2]$. The parameter $h$ determines the amount of smoothing and should be set relative to the number of grid points used to simulate the lattice. We used $h=0.01$ throughout.

\begin{figure}
\includegraphics[width=0.8\linewidth]{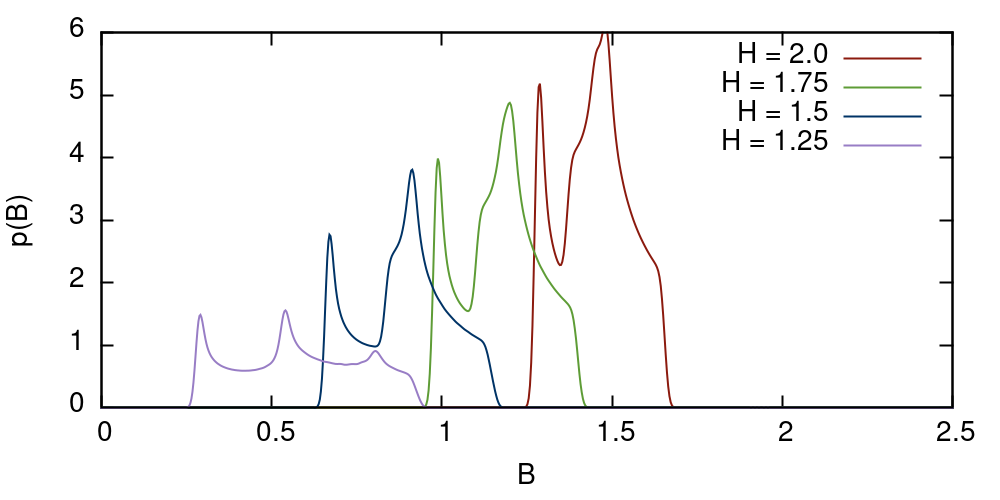}
\includegraphics[width=0.8\linewidth]{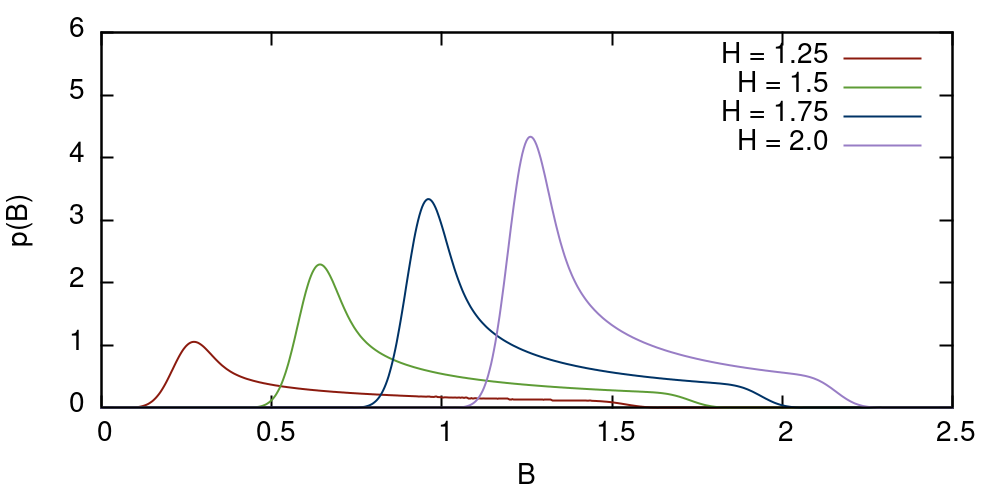}
\caption{\label{fig:nematicProb}
Magnetic field distribution for a skyrmion state in nematic superconductor (top) and vortex lattice in single component superconductor (bottom) for external fields closer to $H_{c1}$ than $H_{c2}$. The nematic superconductor is described in \eqref{eq:G} with $\eta = 3$.  The structure of the field distribution is markedly different for the skyrmion state in the nematic superconductor with a double peak, rather than the familiar single peak for the standard GL model.}
\end{figure}

We used the above method to construct the field distribution $p(B)$ for a nematic superconductor and a standard single component superconductor for comparison. The latter's parameters were chosen so that $H_{c_1}$ and $H_{c_2}$ match those of the nematic model for $\eta = 3$. This process and the parameters are described in the appendix. We have plotted the nematic field distribution for field strengths nearer $H_{c1}$ than $H_{c2}$ in \figref{fig:nematicProb}, and also compared the results to the standard GL model. There is a clear difference, with the nematic superconductor exhibiting a double peak structure as opposed to the well known single peak of the standard GL model. It is also different from the $\mu SR$ signature of a system that phase separates into vortex clusters and domains of Meissner states \cite{biswas2020coexistence,ray2014muon}.

\begin{figure}
\includegraphics[width=0.8\linewidth]{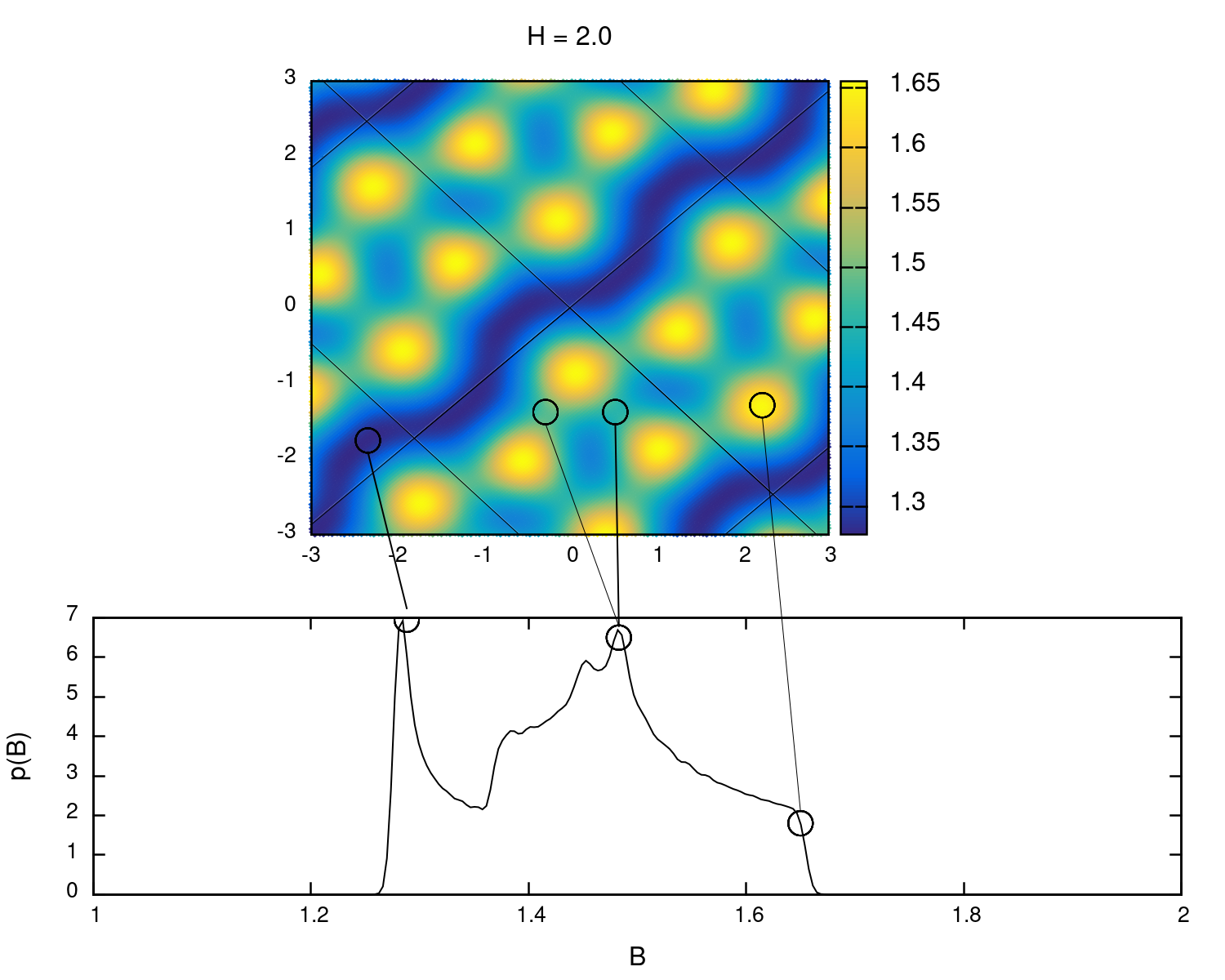}\caption{\label{fig:comparePlot}
Comparing the field distribution $p(B)$ of a nematic superconductor in the lower plot with the areas of the local magnetic field configuration $B(x)$ each peak corresponds to for external field strength $H=2$.}
\end{figure}

\begin{figure}
\includegraphics[width=0.8\linewidth]{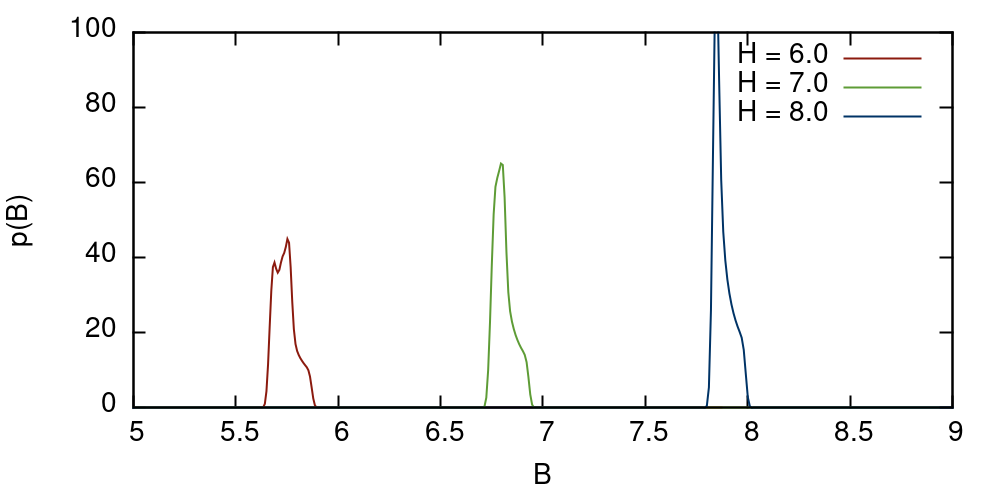}
\includegraphics[width=0.8\linewidth]{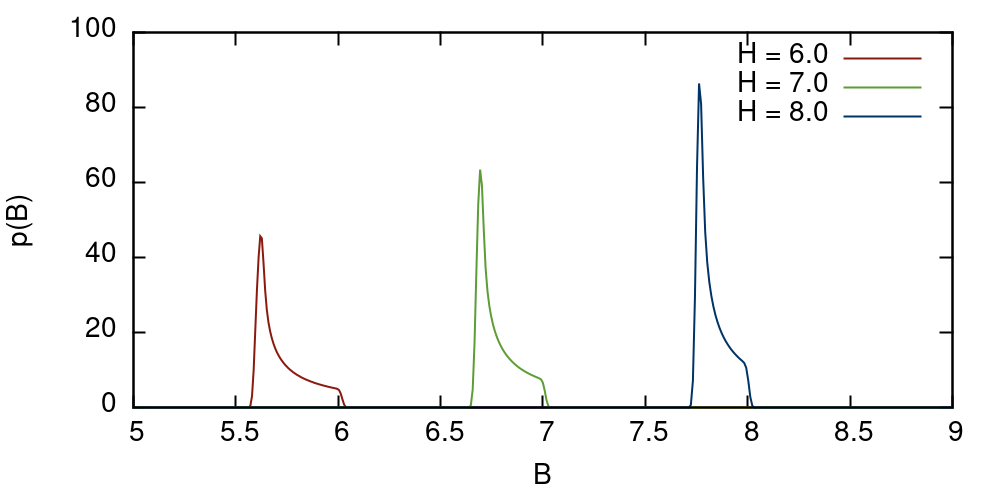}
\caption{\label{fig:nematicProbUpper}
Magnetic field distribution for a nematic superconductor (top) and single component superconductor (bottom) for external fields  $H_{c2} \approx H  \gg H_{c1}$. The nematic superconductor is described in \eqref{eq:G} with $\eta = 3$.  As one approaches $H_{c2}$, the structure of the field distribution for the nematic superconductor  becomes similar to the single peak structure of the familiar single peak for the standard GL model. This corresponds with the lattice becoming a triangular lattice for external fields near $H_{c2}$.}
\end{figure}

Consider the single component Ginzburg-Landau model. Here, the neighbouring vortex separation is determined by a single parameter, the Ginzburg-Landau parameter $\kappa$.  While this parameter is all that is needed to describe the lattice solution, it is actually the ratio of two length scales.  The form of the magnetic field distribution is determined by how the magnetic field decays between neighbouring vortices, hence as there is a single defining parameter, there is a single corresponding peak in the field distribution. In contrast, multicomponent anisotropic models have a continuous set of coupled length scales that are directionally dependent \cite{speight2019chiral,nematic}. However, given the rectangular nematic unit cell shown in \figref{fig:lattice}, we can describe the vortex separation in the unit cell using just two length scales, $|V_1|$ and $|V_2|$. Hence, the decay of the magnetic field between two neighbouring vortices depends on whether they are in the same chain or neighbouring chains. If the vortices are in the same chain the decay is determined by $|V_1|$, if they are in different chains the decay is determined by $|V_2|$.  Hence, each of these neighbouring decay possibilities, or length scales, contributes a different peak to the field distribution $p(B)$.  This can be seen in \figref{fig:comparePlot}, where we have compared the probability density distribution $p(B)$ with the corresponding magnetic field plot $B(x)$ for $H=2$. It can be seen in this figure that the first sharp peak corresponds to chain separation, and the second peak corresponds to the chain link length.

The two peaks for low fields start far apart, due to the large chain separation $|V_2|$. Then as the field strength increases, the chains are packed in much tighter and $|V_2|$ decreases, causing the two peaks to get closer. Then as the external field strength is increased further, both $|V_1|$ and $|V_2|$ shrink. This explains why the peaks begin far apart for particularly low fields. Hence, we ultimately end up with a very distinctive double peak structure for nematic skyrmion lattices. We expect that in the presence of weak pinning the statistics of the magnetic fields in the bulk will still retain these features.

While the above analysis holds for low and medium strength external fields, once the external field is increased to be near $H_{c2}$, the vortices are forced into a tightly packed triangular lattice. This triangular structure leads to the most efficient packing for the vortices, hence it is natural that the increased pressure will eventually force a triangular lattice to be lower in energy. The vortex separation for the nematic superconductor is now described by a single length scale and forms a single peak, similar to that of the single component model. This can be seen for strong external fields in \figref{fig:nematicProbUpper}, where we have again compared the nematic and single component results.

Note that \figref{fig:nematicProb} has a low field peak separation of about $0.5$ which is approximately $0.13 H_{c_2}$ (where $H_{c_2}$ is calculated in \cite{nematic}). Modern muon spectroscopy techniques are capable of resolving peak separations far lower than this in various materials.  Experiments often exhibit a background peak at $H_0$ from muons that overshoot the sample.  This can be resolved from the single type-2 peak despite being far closer than $10 \%$ of $H_{c_2}$ \cite{sonier2000musr}.

In conclusion we have shown that the magnetic response of nematic superconductors takes the form of skyrmion stripes, which is markedly different from the standard vortex lattice. As the field is increased, these skyrmion stripes decrease their separation. Eventually the magnetization process compresses the skyrmion stripes themselves, forming the conventional triangular vortex lattice close to upper critical field $H_{c2}$. We also find that skyrmion stripes give qualitatively distinct signatures in muon spin rotation experiments. In particular, we observe a distinctive two peak structure, where the second peak changes relative to the first as the external field strength is increased. These findings can be used to experimentally confirm (i) nematic superconducting states and (ii) that flux-carrying objects in superconductors take the form of skyrmions. We have also presented new numerical tools, to find unit cells of any symmetry, that can be used to analyse the signals of other unconventional materials.

\section*{Acknowledgements}
We thank Vadim Grinenko and Andrew Huxley for useful discussions. EB was supported by the Swedish Research Council Grants 2016-06122, 2018-03659, and Olle Engkvists Stiftelse. MS and TW were supported by the UK Engineering and Physical Sciences Research Council, through grant EP/P024688/1. TW was also supported by an academic development fellowship, awarded by the University of Leeds, where he was based for the majority of the work. The numerical work of this paper was performed using the code library Soliton Solver, developed by one of the authors, and was undertaken on ARC4, part of the High Performance Computing facilities at the University of Leeds.

\newpage
\appendix
\section{Finding vortex lattice solutions}
In this section we give a brief account of the new method used to find  vortex lattice solutions of arbitrary complexity in the presence of an external field $H$. Detailed discussion of the method is presented in the companion paper \cite{nematic}. Namely, we find the unit cell of the periodic vortex solution in the bulk, with external field strength $H_{c_1} < H_0 < H_{c_2}$. The standard approach for a Ginzburg-Landau model is to consider a unit cell of degree $n=1$ with either triangular ($\alpha = \pi/3$) or square ($\alpha = \pi/2$) symmetry. However, as the nematic system we are considering is anisotropic,  there is no reason to expect that a lattice with such high symmetry will be the global minimizer. Hence, the correct approach is to minimize energy, not just w.r.t.\ the periodic fields, but also w.r.t.\ the geometry of the unit cell itself. We present here a general method to find the optimal unit cell, without assuming the symmetry of the underlying lattice, based upon the method first presented in \cite{nematic}.

We first assume that far from the boundary of the system (deep in the bulk), the gauge invariant quantities $(\rho_\alpha,\theta_{12},B)$ are periodic on the unit cell.  We also note that we can represent a general periodic structure as a tessellation of a general parallelogram (unit cell), as seen in figure \ref{fig:unitCell}, formed by two vectors $V_1$ and $V_2$ with angle $\alpha$.

We then seek local minimisers of the Gibbs free energy in \eqref{eq:G} w.r.t. the fields $\psi_\alpha(x), A_i(x)$, now defined on a general flat periodic 2-torus $T_\Lambda^2 = \mathbb{R}^2/\Lambda$, where the period lattice $\Lambda$ is spanned by vectors $V_1$, $V_2$ in \figref{fig:unitCell},
\begin{equation}
\Lambda = \left\{ n_1V_1 + n_2 V_2 \,|\, n_1, n_2 \in \mathbb{Z}, V_1, V_2 \in \mathbb{R}^2\right\}.
\end{equation}
The unit cell has periodic boundary conditions, up to gauge, so that the fields have boundary conditions,
\begin{align}
\psi_\alpha( x + V_i ) &= \psi_\alpha (x)e^{i f_i},\nonumber\\
A_j ( x + V_i ) &=  A_j(x) + \partial_j f_i
\end{align}
Note that while the fields are not purely periodic, the above boundary conditions do leave all physical quantities periodic as desired,
\begin{align}
B(x + V_i) = B(x), \quad \,J(x + V_i) = J(x),\nonumber\\ |\psi(x + V_i)| = |\psi(x)|. \quad\quad\quad\quad
\end{align}

To simulate the fields on the unit cell, we will simplify the above formulation by transforming to a more convenient coordinate system in the $x, y$ plane. Let us define $X_1,X_2$ so that $(x,y)=X_1V_1+X_2V_2$. The unit cell spanned by $V_1,V_2$ is now covered by $(X_1,X_2)\in[0,1]\times[0,1]$. Let $L$ be the matrix with columns $V_1, V_2$. Note that $\det L=\area$, the area of the unit cell. Now define the unimodular matrix $M=\sqrt\area L^{-1}$ such that $X_i = \frac{M_{ij}}{\sqrt{\area}} x_j$. Then
\begin{widetext}
\begin{align*}
F(M,\area,\psi_\alpha,A) &= \int_{[0,1]^2} \left\{\frac{1}{2} (M Q^{\alpha\beta} M^T)_{ij}\overline{D_i \psi_\alpha} D_j \psi_\beta + \frac{1}{2\area} \left((\partial_1 A_2 - \partial_2 A_1)^2 \right)
+\area F_p(\psi_\alpha)\right\} dX_1 \, dX_2,
\label{eq:Esq} 
\end{align*}
\end{widetext}
leading in turn to the Gibbs free energy,
\begin{align}
\nonumber G(M,\area,\psi_\alpha,A) &= F - \int_{[0,1]^2} H_0(\partial_1 A_2 - \partial_2 A_1)\, dX_1 \, dX_2 \\
&= F - 2n\pi H_0.
\end{align}
where $n$ is the winding number of the field configuration. Note that this is only valid in the basal plane, as otherwise one must consider non-zero $A_3$ in the free energy. This would allow in-plane spontaneous fields and magnetic field twisting, which occurs away from the basal plane due to the coupled length scales\cite{nematic}. Models that exhibit such fields have been studied in \cite{speight2021magnetic,benfenati2020magnetic,nematic}.

If we exclude boundary effects, the optimal lattice solution is the one that minimises the the total Gibbs free energy of the system. If the system area is $\area_{sys}$, this can be calculated from the unit cell as $G_{sys} = \area_{sys}G/\area$. Hence we seek minimisers of $G/\area$ or Gibbs free energy per unit area, with respect to the fields, geometry of the unit cell $M$ and area $\area$. Note that the degree or winding number of a given unit cell $n$ is fixed and strictly speaking we must the find the global minimiser of all $G_n/\area$, where $G_n$ is the Gibbs free energy of a unit cell of degree $n$ and then find the infimum of this set. 

In practice, it is sufficient to find minima of $G_n/\area$ until we get a repeated minimiser, that is, until we find a minimiser of $G_n/\area$ whose cell and field configuration is two cells of the $G_{n/2}/\area$ minimiser joined together.

\subsection{Numerical method}

To find minimisers of $G(M,\area,\psi_\alpha,A)$ we discretise the standard square unit cell on an $n_1 \times n_2$ regular grid. We utilize an arrested Newton flow algorithm subject to the periodic boundary conditions,
\begin{align}
\psi_\alpha( x + (1,0) ) &= \psi_\alpha (x)e^{i n2\pi n x_2},\nonumber\\
\psi_\alpha( x + (0,1) ) &= \psi_\alpha (x),\nonumber\\
A_2 ( x + (1,0) ) &= A_1(x) + 2\pi n \nonumber \\
A_i ( x + (1,0) ) &= A_2(x)\quad i \neq 2 \nonumber \\
A_i ( x + (0,1) ) &= A_i(x)
\end{align}
where $n$ is the winding number of the unit cell. We set the fields $\phi = (\psi, A)$ on the unit square torus and the geometry of the unit cell $(M,\area)$ to be some initial condition. Fixing the unit cell $(M,\area)$, we find a local minimum w.r.t. the collected fields $\phi$, using arrested Newton flow for a notional particle subject to the potential $\ddot{\phi} = - grad \, G_{dis}(M,\area,\phi)$. A more detailed description of arrested newton flow is given in \cite{speight2020skyrmions}. This is continued until the discretised function is minimised with respect to some tolerance. We then fix the field configuration $\phi$ and area $\area$, and minimise $G_{dis}(M,\area,\phi)/\area$ with respect to $M \in SL(2,\mathbb{R})$, where we note that only the gradient term of the energy is dependent on this matrix. This can be done exactly using linear algebra, described in the next subsection. Finally, we fix the fields $\phi$ and the lattice shape $M$ and minimise the free energy w.r.t. $\area$, by solving the algebraic equation in $\area$ with coefficients given by the integrals of the different energy terms. 

\subsection{Finding the minimal geometry}
We can find the minimal geometry of a unit cell with a given configuration $(\psi_\alpha, A)$ explicitly. We first note that the only term that is dependent on the geometry $M$ is the gradient term of the free energy. Hence we write the gradient energy as a $4 \times 4$ complex matrix,
\begin{equation}
P^c_{ik,jl} := \int_{[0,1]^2}Q^{\alpha\beta}_{kl} \overline{D_i \psi_\alpha}D_j \psi_\beta,
\end{equation}
where we take each index pair $(i,k), (j,l)$ to range over the values $\left\{(1,1),(1,2),(2,1),(2,2)\right\}$. Due to the symmetry $Q^{\alpha\beta}_{ij} = \ol{Q^{\beta\alpha}_{ji}}$ we know $P^c$ is hermitian. Hence if we split it into its real and imaginary parts $P^c = P + i P_I$ we know that the real part $P$ is symmetric and the imaginary part $P_I$ is skew. If we then consider the vector,
\begin{equation}
X := \left( M_{11} M_{12} M_{21} M_{22} \right)^T,
\end{equation}
we can rewrite the gradient part of the free energy as,
\begin{equation}
F_{grad}(X) = \frac{1}{2} X^T P^c X = \frac{1}{2} X^T P X.
\end{equation}
Hence we must minimise the quadratic form $X^T P X$ subject to the constraint that $\det M = 1$ (or $M \in SL(2,\mathbb{R}$), which can be written,
\begin{equation}
\frac{1}{2} X^T J X = 1, \quad \quad J := \left( \begin{array}{cccc} 0 & 0 & 0 & 1\\ 0 & 0 & -1 & 0 \\ 0 & -1 & 0 & 0 \\ 1 & 0 & 0 & 0 \end{array}\right).
\end{equation}
Hence, we seek critical points of $F_{\text{grad}}:SL(2,\mathbb{R}) \rightarrow \mathbb{R}$, which are $X \in \mathbb{R}^4$ with $X^T J X = 1$ and,
\begin{equation}
P X = \lambda \left(X_4, -X_3, -X_2, X_1\right)^T = \lambda J X,
\label{eq:linearEq}
\end{equation}
for some $\lambda \in \mathbb{R}$. We can see that $J^2 = \mathbb{I}_4$ is the identity matrix, hence \eqref{eq:linearEq} becomes,
\begin{equation}
J P X = \lambda X.
\end{equation}
In other words $X$ is an eigenvector of $J P$. 

This allows us to minimise $F_{\text{grad}}$ w.r.t. $M$ explicitly by constructing $J P$ and finding it's 4 eigenvectors $Y$. We discard any complex solutions and any s.t. $Y^T J Y \leq 0$. We compute the solution we seek $X = Y/\sqrt{ Y^T J Y }$ such that,
\begin{equation}
F_{\text{grad}} = \frac{1}{2} X^T P X = \frac{1}{2} X^T \lambda J X = \lambda.
\end{equation}
We can then read off the new period lattice as,
\begin{equation}
L = \sqrt{\area}M^{-1}, \quad \quad M = \left( \begin{array}{cc} X_1 & X_2 \\ X_3 & X_4 \end{array}\right).
\end{equation}
The period lattice is spanned by the columns of $L$, named $V_1$, $V_2$. 

\section{Single component field distribution}
It is instructive to compare the results for nematic systems to a standard single component Ginzburg Landau model.  We will consider the following model,
\begin{equation}
G = \int_{\mathbb{R}^2}\left\{\frac{1}{2}\ol{D_i \psi} D_i \psi + \frac{1}{2} |B-H|^2 + F_p(|\psi|)\right\},
\label{eq:Gsingle}
\end{equation}
where there is a single order parameter $\psi$, and
\begin{equation}
F_p(|\psi|) = - a|\psi|^2 + \frac{b}{2} |\psi|^4.
\end{equation}

To effectively compare this model to the results for the nematic model, we will select the parameters $a,b$ so that $H_{c_1}$ and $H_{c_2}$ equal those of the nematic model. By a standard computation, for example see Tinkham page 134 \cite{tinkham2004introduction},
\begin{equation}
H_{c_2} = 2a,
\end{equation}
while, by definition, $H_{c1}$ is the value of external field strength $H$ for which the Gibbs energy of a vortex solution equals that of the superconducting ground state, yielding
\begin{equation}\label{hc1def}
H_{c1}=\frac{F_{\scriptscriptstyle vortex}-
F_{\scriptscriptstyle ground}}{2\pi}.
\end{equation}
The right hand side of \eqref{hc1def} depends implicitly on $a,b$. We find that $a=5.70$, $b=2.21$ produces $H_{c1}=1.18$ and $H_{c2}=11.41$, matching the nematic model with $\eta=3$.

\section{Results for $\eta = 10$}

\begin{figure*}
\includegraphics[width=1.0\linewidth]{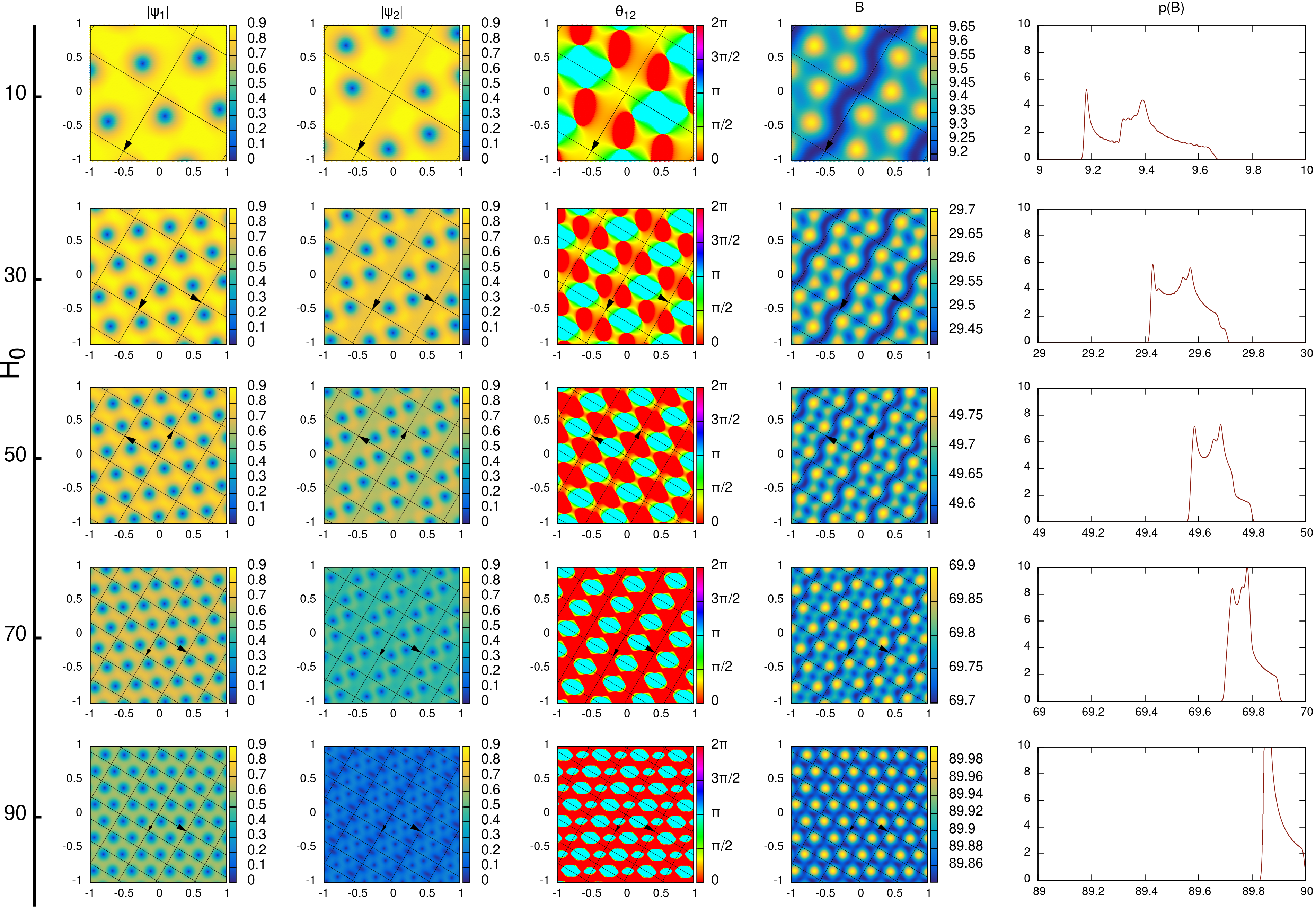}\caption{\label{fig:lattice_eta10}
Contour plots of the lattice solutions (skyrmion chains) for the model \eqref{eq:G} in the basal plane, with $\eta = 10$ and increasing external field strength $H_0$. The fields plotted are the order parameter magnitudes $|\psi_1|$ and $|\psi_2|$, the phase difference between the two complex order parameters $\theta_{12} = \varphi_1 - \varphi_2$ and the local magnetic field strength $B$.  The final column gives the magnetic field probability density $p(B)$ for the corresponding lattice solution.  Similar to other results in this work,  the low field solutions take the form of skyrmion chains, causing $p(B)$ to have a double peak structure. Then as the external field increases the skyrmion chains squash together causing the two peaks in $p(B)$ to get closer. Then near $H_{c_2}$ a transition occurs and the vortices are forced into the conventional triangular lattice and the double peak becomes a single peak. }
\end{figure*}

It should be emphasised that the behaviour discussed in the body of the paper is not fine tuned by the parameters and exists for a wide choice for $\eta$. As $\eta$ controls the ratio of $H_{c_2}$ and $H_{c_1}$ this is a vital parameter for fitting the behaviour to a particular material where this ratio is known.  The choice of $\eta = 3$ in \figref{fig:lattice} and \figref{fig:nematicProb} corresponds to a ratio of $H_{c_2}(\hat{z})/H_{c_1}(\hat{z}) \approx 9$, however it has been suggested that some nematic materials have a ratio an order of magnitude higher than this. To this end we provide additional results for $\eta = 10$ in \figref{fig:lattice_eta10}, where the critical field ratio is $H_{c_2}(\hat{z}) / H_{c_1}(\hat{z}) \approx 100$. 

The results in \figref{fig:lattice_eta10} for $\eta = 10$ demonstrate the same qualitative properties as for $\eta = 3$.  In particular, near $H_{c_1}$ we observe  skyrmion chains with a rectangular unit cell.  The resulting magnetic distribution has a similar double peak structure.  Note that as $\eta$ scales the potential the vortex size is smaller decreasing both the chain link length $V_1$ and the chain separation $V_2$ accordingly.  As the strength of the external field $|H|$ is increased, the chain separation decreases. This corresponds with the separation of the magnetic field probability peaks decreasing. Finally, as the two peaks overlap the lattice becomes a triangular lattice of composite vortices and the skyrmions disappear.

\section{Relation to current experimental surface probes of $Cu_xBi_2Se_3$}
It is worth noting that vortex lattices have been studied in $Cu_xBi_2Se_3$ using scanning tunneling microscopy (STM)\cite{tao2018direct} and skyrmion stripes were not observed. This is not necessarily incompatible with the results presented here, as STM measures surface states, whereas we explicitly consider only the bulk superconducting states. It has been speculated that such exotic superconductors exhibit different vortex properties and even different order parameters on the surface and in the bulk\cite{tao2018direct}. In addition, the effects of stray fields and pinning are stronger on the surface than in the bulk. This highlights the importance of our proposal of extracting information about nematic superconductivity from the bulk properties of vortices in various materials using $\mu SR$.
\bibliography{bibliography}

\begin{thebibliography}{46}
\expandafter\ifx\csname natexlab\endcsname\relax\def\natexlab#1{#1}\fi
\expandafter\ifx\csname bibnamefont\endcsname\relax
  \def\bibnamefont#1{#1}\fi
\expandafter\ifx\csname bibfnamefont\endcsname\relax
  \def\bibfnamefont#1{#1}\fi
\expandafter\ifx\csname citenamefont\endcsname\relax
  \def\citenamefont#1{#1}\fi
\expandafter\ifx\csname url\endcsname\relax
  \def\url#1{\texttt{#1}}\fi
\expandafter\ifx\csname urlprefix\endcsname\relax\def\urlprefix{URL }\fi
\providecommand{\bibinfo}[2]{#2}
\providecommand{\eprint}[2][]{\url{#2}}

\bibitem[{\citenamefont{Fu and Berg}(2010)}]{fu2010odd}
\bibinfo{author}{\bibfnamefont{L.}~\bibnamefont{Fu}} \bibnamefont{and}
  \bibinfo{author}{\bibfnamefont{E.}~\bibnamefont{Berg}},
  \bibinfo{journal}{Physical review letters} \textbf{\bibinfo{volume}{105}},
  \bibinfo{pages}{097001} (\bibinfo{year}{2010}).

\bibitem[{\citenamefont{Fu}(2014)}]{fu2014odd}
\bibinfo{author}{\bibfnamefont{L.}~\bibnamefont{Fu}},
  \bibinfo{journal}{Physical Review B} \textbf{\bibinfo{volume}{90}},
  \bibinfo{pages}{100509} (\bibinfo{year}{2014}).

\bibitem[{\citenamefont{Venderbos
  et~al.}(2016{\natexlab{a}})\citenamefont{Venderbos, Kozii, and
  Fu}}]{venderbos2016odd}
\bibinfo{author}{\bibfnamefont{J.~W.} \bibnamefont{Venderbos}},
  \bibinfo{author}{\bibfnamefont{V.}~\bibnamefont{Kozii}}, \bibnamefont{and}
  \bibinfo{author}{\bibfnamefont{L.}~\bibnamefont{Fu}},
  \bibinfo{journal}{Physical Review B} \textbf{\bibinfo{volume}{94}},
  \bibinfo{pages}{180504} (\bibinfo{year}{2016}{\natexlab{a}}).

\bibitem[{\citenamefont{Wang et~al.}(2021)\citenamefont{Wang, Wu, Zhou, Li,
  Teng, Dong, He, Zhang, Ding, and Li}}]{wang2021progress}
\bibinfo{author}{\bibfnamefont{J.}~\bibnamefont{Wang}},
  \bibinfo{author}{\bibfnamefont{Y.}~\bibnamefont{Wu}},
  \bibinfo{author}{\bibfnamefont{X.}~\bibnamefont{Zhou}},
  \bibinfo{author}{\bibfnamefont{Y.}~\bibnamefont{Li}},
  \bibinfo{author}{\bibfnamefont{B.}~\bibnamefont{Teng}},
  \bibinfo{author}{\bibfnamefont{P.}~\bibnamefont{Dong}},
  \bibinfo{author}{\bibfnamefont{J.}~\bibnamefont{He}},
  \bibinfo{author}{\bibfnamefont{Y.}~\bibnamefont{Zhang}},
  \bibinfo{author}{\bibfnamefont{Y.}~\bibnamefont{Ding}}, \bibnamefont{and}
  \bibinfo{author}{\bibfnamefont{J.}~\bibnamefont{Li}},
  \bibinfo{journal}{Advances in Physics: X} \textbf{\bibinfo{volume}{6}},
  \bibinfo{pages}{1878931} (\bibinfo{year}{2021}).

\bibitem[{\citenamefont{Zyuzin et~al.}(2017)\citenamefont{Zyuzin, Garaud, and
  Babaev}}]{zyuzin2017nematic}
\bibinfo{author}{\bibfnamefont{A.}~\bibnamefont{Zyuzin}},
  \bibinfo{author}{\bibfnamefont{J.}~\bibnamefont{Garaud}}, \bibnamefont{and}
  \bibinfo{author}{\bibfnamefont{E.}~\bibnamefont{Babaev}},
  \bibinfo{journal}{Physical review letters} \textbf{\bibinfo{volume}{119}},
  \bibinfo{pages}{167001} (\bibinfo{year}{2017}).

\bibitem[{\citenamefont{Wu and Martin}(2017)}]{wu2017majorana}
\bibinfo{author}{\bibfnamefont{F.}~\bibnamefont{Wu}} \bibnamefont{and}
  \bibinfo{author}{\bibfnamefont{I.}~\bibnamefont{Martin}},
  \bibinfo{journal}{Physical Review B} \textbf{\bibinfo{volume}{95}},
  \bibinfo{pages}{224503} (\bibinfo{year}{2017}).

\bibitem[{\citenamefont{Chirolli}(2018)}]{chirolli2018chiral}
\bibinfo{author}{\bibfnamefont{L.}~\bibnamefont{Chirolli}},
  \bibinfo{journal}{Physical Review B} \textbf{\bibinfo{volume}{98}},
  \bibinfo{pages}{014505} (\bibinfo{year}{2018}).

\bibitem[{\citenamefont{Barkman et~al.}(2019)\citenamefont{Barkman, Zyuzin, and
  Babaev}}]{barkman2019antichiral}
\bibinfo{author}{\bibfnamefont{M.}~\bibnamefont{Barkman}},
  \bibinfo{author}{\bibfnamefont{A.~A.} \bibnamefont{Zyuzin}},
  \bibnamefont{and} \bibinfo{author}{\bibfnamefont{E.}~\bibnamefont{Babaev}},
  \bibinfo{journal}{Physical Review B} \textbf{\bibinfo{volume}{99}},
  \bibinfo{pages}{220508} (\bibinfo{year}{2019}).

\bibitem[{\citenamefont{Uematsu et~al.}(2019)\citenamefont{Uematsu, Mizushima,
  Tsuruta, Fujimoto, and Sauls}}]{uematsu2019chiral}
\bibinfo{author}{\bibfnamefont{H.}~\bibnamefont{Uematsu}},
  \bibinfo{author}{\bibfnamefont{T.}~\bibnamefont{Mizushima}},
  \bibinfo{author}{\bibfnamefont{A.}~\bibnamefont{Tsuruta}},
  \bibinfo{author}{\bibfnamefont{S.}~\bibnamefont{Fujimoto}}, \bibnamefont{and}
  \bibinfo{author}{\bibfnamefont{J.}~\bibnamefont{Sauls}},
  \bibinfo{journal}{Physical review letters} \textbf{\bibinfo{volume}{123}},
  \bibinfo{pages}{237001} (\bibinfo{year}{2019}).

\bibitem[{\citenamefont{Dentelski et~al.}(2020)\citenamefont{Dentelski, Kozii,
  and Ruhman}}]{dentelski2020effect}
\bibinfo{author}{\bibfnamefont{D.}~\bibnamefont{Dentelski}},
  \bibinfo{author}{\bibfnamefont{V.}~\bibnamefont{Kozii}}, \bibnamefont{and}
  \bibinfo{author}{\bibfnamefont{J.}~\bibnamefont{Ruhman}},
  \bibinfo{journal}{Physical Review Research} \textbf{\bibinfo{volume}{2}},
  \bibinfo{pages}{033302} (\bibinfo{year}{2020}).

\bibitem[{\citenamefont{How and Yip}(2020{\natexlab{a}})}]{yiphalf}
\bibinfo{author}{\bibfnamefont{P.~T.} \bibnamefont{How}} \bibnamefont{and}
  \bibinfo{author}{\bibfnamefont{S.-K.} \bibnamefont{Yip}},
  \bibinfo{journal}{Phys. Rev. Research} \textbf{\bibinfo{volume}{2}},
  \bibinfo{pages}{043192} (\bibinfo{year}{2020}{\natexlab{a}}),
  \urlprefix\url{https://link.aps.org/doi/10.1103/PhysRevResearch.2.043192}.

\bibitem[{\citenamefont{Chichinadze et~al.}(2020)\citenamefont{Chichinadze,
  Classen, and Chubukov}}]{chubukovTBG}
\bibinfo{author}{\bibfnamefont{D.~V.} \bibnamefont{Chichinadze}},
  \bibinfo{author}{\bibfnamefont{L.}~\bibnamefont{Classen}}, \bibnamefont{and}
  \bibinfo{author}{\bibfnamefont{A.~V.} \bibnamefont{Chubukov}},
  \bibinfo{journal}{Phys. Rev. B} \textbf{\bibinfo{volume}{102}},
  \bibinfo{pages}{125120} (\bibinfo{year}{2020}),
  \urlprefix\url{https://link.aps.org/doi/10.1103/PhysRevB.102.125120}.

\bibitem[{\citenamefont{Cho et~al.}(2020)\citenamefont{Cho, Shen, Lyu, Atanov,
  Chen, Lee, Hor, Gawryluk, Pomjakushina, Bartkowiak et~al.}}]{cho2020z3}
\bibinfo{author}{\bibfnamefont{C.-w.} \bibnamefont{Cho}},
  \bibinfo{author}{\bibfnamefont{J.}~\bibnamefont{Shen}},
  \bibinfo{author}{\bibfnamefont{J.}~\bibnamefont{Lyu}},
  \bibinfo{author}{\bibfnamefont{O.}~\bibnamefont{Atanov}},
  \bibinfo{author}{\bibfnamefont{Q.}~\bibnamefont{Chen}},
  \bibinfo{author}{\bibfnamefont{S.~H.} \bibnamefont{Lee}},
  \bibinfo{author}{\bibfnamefont{Y.~S.} \bibnamefont{Hor}},
  \bibinfo{author}{\bibfnamefont{D.~J.} \bibnamefont{Gawryluk}},
  \bibinfo{author}{\bibfnamefont{E.}~\bibnamefont{Pomjakushina}},
  \bibinfo{author}{\bibfnamefont{M.}~\bibnamefont{Bartkowiak}},
  \bibnamefont{et~al.}, \bibinfo{journal}{Nature communications}
  \textbf{\bibinfo{volume}{11}}, \bibinfo{pages}{1} (\bibinfo{year}{2020}).

\bibitem[{\citenamefont{Hinkov et~al.}(2008)\citenamefont{Hinkov, Haug,
  Fauqu{\'e}, Bourges, Sidis, Ivanov, Bernhard, Lin, and
  Keimer}}]{hinkov2008electronic}
\bibinfo{author}{\bibfnamefont{V.}~\bibnamefont{Hinkov}},
  \bibinfo{author}{\bibfnamefont{D.}~\bibnamefont{Haug}},
  \bibinfo{author}{\bibfnamefont{B.}~\bibnamefont{Fauqu{\'e}}},
  \bibinfo{author}{\bibfnamefont{P.}~\bibnamefont{Bourges}},
  \bibinfo{author}{\bibfnamefont{Y.}~\bibnamefont{Sidis}},
  \bibinfo{author}{\bibfnamefont{A.}~\bibnamefont{Ivanov}},
  \bibinfo{author}{\bibfnamefont{C.}~\bibnamefont{Bernhard}},
  \bibinfo{author}{\bibfnamefont{C.}~\bibnamefont{Lin}}, \bibnamefont{and}
  \bibinfo{author}{\bibfnamefont{B.}~\bibnamefont{Keimer}},
  \bibinfo{journal}{Science} \textbf{\bibinfo{volume}{319}},
  \bibinfo{pages}{597} (\bibinfo{year}{2008}).

\bibitem[{\citenamefont{Daou et~al.}(2010)\citenamefont{Daou, Chang, LeBoeuf,
  Cyr-Choiniere, Lalibert{\'e}, Doiron-Leyraud, Ramshaw, Liang, Bonn, Hardy
  et~al.}}]{daou2010broken}
\bibinfo{author}{\bibfnamefont{R.}~\bibnamefont{Daou}},
  \bibinfo{author}{\bibfnamefont{J.}~\bibnamefont{Chang}},
  \bibinfo{author}{\bibfnamefont{D.}~\bibnamefont{LeBoeuf}},
  \bibinfo{author}{\bibfnamefont{O.}~\bibnamefont{Cyr-Choiniere}},
  \bibinfo{author}{\bibfnamefont{F.}~\bibnamefont{Lalibert{\'e}}},
  \bibinfo{author}{\bibfnamefont{N.}~\bibnamefont{Doiron-Leyraud}},
  \bibinfo{author}{\bibfnamefont{B.}~\bibnamefont{Ramshaw}},
  \bibinfo{author}{\bibfnamefont{R.}~\bibnamefont{Liang}},
  \bibinfo{author}{\bibfnamefont{D.}~\bibnamefont{Bonn}},
  \bibinfo{author}{\bibfnamefont{W.}~\bibnamefont{Hardy}},
  \bibnamefont{et~al.}, \bibinfo{journal}{Nature}
  \textbf{\bibinfo{volume}{463}}, \bibinfo{pages}{519} (\bibinfo{year}{2010}).

\bibitem[{\citenamefont{Ronning et~al.}(2017)\citenamefont{Ronning, Helm,
  Shirer, Bachmann, Balicas, Chan, Ramshaw, Mcdonald, Balakirev, Jaime
  et~al.}}]{ronning2017electronic}
\bibinfo{author}{\bibfnamefont{F.}~\bibnamefont{Ronning}},
  \bibinfo{author}{\bibfnamefont{T.}~\bibnamefont{Helm}},
  \bibinfo{author}{\bibfnamefont{K.}~\bibnamefont{Shirer}},
  \bibinfo{author}{\bibfnamefont{M.}~\bibnamefont{Bachmann}},
  \bibinfo{author}{\bibfnamefont{L.}~\bibnamefont{Balicas}},
  \bibinfo{author}{\bibfnamefont{M.~K.} \bibnamefont{Chan}},
  \bibinfo{author}{\bibfnamefont{B.}~\bibnamefont{Ramshaw}},
  \bibinfo{author}{\bibfnamefont{R.~D.} \bibnamefont{Mcdonald}},
  \bibinfo{author}{\bibfnamefont{F.~F.} \bibnamefont{Balakirev}},
  \bibinfo{author}{\bibfnamefont{M.}~\bibnamefont{Jaime}},
  \bibnamefont{et~al.}, \bibinfo{journal}{Nature}
  \textbf{\bibinfo{volume}{548}}, \bibinfo{pages}{313} (\bibinfo{year}{2017}).

\bibitem[{\citenamefont{Okazaki et~al.}(2011)\citenamefont{Okazaki, Shibauchi,
  Shi, Haga, Matsuda, Yamamoto, Onuki, Ikeda, and
  Matsuda}}]{okazaki2011rotational}
\bibinfo{author}{\bibfnamefont{R.}~\bibnamefont{Okazaki}},
  \bibinfo{author}{\bibfnamefont{T.}~\bibnamefont{Shibauchi}},
  \bibinfo{author}{\bibfnamefont{H.}~\bibnamefont{Shi}},
  \bibinfo{author}{\bibfnamefont{Y.}~\bibnamefont{Haga}},
  \bibinfo{author}{\bibfnamefont{T.}~\bibnamefont{Matsuda}},
  \bibinfo{author}{\bibfnamefont{E.}~\bibnamefont{Yamamoto}},
  \bibinfo{author}{\bibfnamefont{Y.}~\bibnamefont{Onuki}},
  \bibinfo{author}{\bibfnamefont{H.}~\bibnamefont{Ikeda}}, \bibnamefont{and}
  \bibinfo{author}{\bibfnamefont{Y.}~\bibnamefont{Matsuda}},
  \bibinfo{journal}{Science} \textbf{\bibinfo{volume}{331}},
  \bibinfo{pages}{439} (\bibinfo{year}{2011}).

\bibitem[{\citenamefont{Fernandes et~al.}(2022)\citenamefont{Fernandes, Coldea,
  Ding, Fisher, Hirschfeld, and Kotliar}}]{fernandes2022iron}
\bibinfo{author}{\bibfnamefont{R.~M.} \bibnamefont{Fernandes}},
  \bibinfo{author}{\bibfnamefont{A.~I.} \bibnamefont{Coldea}},
  \bibinfo{author}{\bibfnamefont{H.}~\bibnamefont{Ding}},
  \bibinfo{author}{\bibfnamefont{I.~R.} \bibnamefont{Fisher}},
  \bibinfo{author}{\bibfnamefont{P.}~\bibnamefont{Hirschfeld}},
  \bibnamefont{and} \bibinfo{author}{\bibfnamefont{G.}~\bibnamefont{Kotliar}},
  \bibinfo{journal}{Nature} \textbf{\bibinfo{volume}{601}}, \bibinfo{pages}{35}
  (\bibinfo{year}{2022}).

\bibitem[{\citenamefont{Wang et~al.}(2023)\citenamefont{Wang, B{\"o}hmer, and
  Fanfarillo}}]{wang2023nematicity}
\bibinfo{author}{\bibfnamefont{Q.}~\bibnamefont{Wang}},
  \bibinfo{author}{\bibfnamefont{A.}~\bibnamefont{B{\"o}hmer}},
  \bibnamefont{and}
  \bibinfo{author}{\bibfnamefont{L.}~\bibnamefont{Fanfarillo}},
  \emph{\bibinfo{title}{Nematicity in iron-based superconductors}}, Frontiers
  Research Topics (\bibinfo{publisher}{Frontiers Media SA},
  \bibinfo{year}{2023}), ISBN \bibinfo{isbn}{9782832514702},
  \urlprefix\url{https://books.google.co.uk/books?id=FxCuEAAAQBAJ}.

\bibitem[{\citenamefont{Chu et~al.}(2012)\citenamefont{Chu, Kuo, Analytis, and
  Fisher}}]{chu2012divergent}
\bibinfo{author}{\bibfnamefont{J.-H.} \bibnamefont{Chu}},
  \bibinfo{author}{\bibfnamefont{H.-H.} \bibnamefont{Kuo}},
  \bibinfo{author}{\bibfnamefont{J.~G.} \bibnamefont{Analytis}},
  \bibnamefont{and} \bibinfo{author}{\bibfnamefont{I.~R.}
  \bibnamefont{Fisher}}, \bibinfo{journal}{Science}
  \textbf{\bibinfo{volume}{337}}, \bibinfo{pages}{710} (\bibinfo{year}{2012}).

\bibitem[{\citenamefont{Lu et~al.}(2014)\citenamefont{Lu, Park, Zhang, Luo,
  Nevidomskyy, Si, and Dai}}]{lu2014nematic}
\bibinfo{author}{\bibfnamefont{X.}~\bibnamefont{Lu}},
  \bibinfo{author}{\bibfnamefont{J.}~\bibnamefont{Park}},
  \bibinfo{author}{\bibfnamefont{R.}~\bibnamefont{Zhang}},
  \bibinfo{author}{\bibfnamefont{H.}~\bibnamefont{Luo}},
  \bibinfo{author}{\bibfnamefont{A.~H.} \bibnamefont{Nevidomskyy}},
  \bibinfo{author}{\bibfnamefont{Q.}~\bibnamefont{Si}}, \bibnamefont{and}
  \bibinfo{author}{\bibfnamefont{P.}~\bibnamefont{Dai}},
  \bibinfo{journal}{Science} \textbf{\bibinfo{volume}{345}},
  \bibinfo{pages}{657} (\bibinfo{year}{2014}).

\bibitem[{\citenamefont{Kuo et~al.}(2016)\citenamefont{Kuo, Chu, Palmstrom,
  Kivelson, and Fisher}}]{kuo2016ubiquitous}
\bibinfo{author}{\bibfnamefont{H.-H.} \bibnamefont{Kuo}},
  \bibinfo{author}{\bibfnamefont{J.-H.} \bibnamefont{Chu}},
  \bibinfo{author}{\bibfnamefont{J.~C.} \bibnamefont{Palmstrom}},
  \bibinfo{author}{\bibfnamefont{S.~A.} \bibnamefont{Kivelson}},
  \bibnamefont{and} \bibinfo{author}{\bibfnamefont{I.~R.}
  \bibnamefont{Fisher}}, \bibinfo{journal}{Science}
  \textbf{\bibinfo{volume}{352}}, \bibinfo{pages}{958} (\bibinfo{year}{2016}).

\bibitem[{\citenamefont{Hor et~al.}(2010)\citenamefont{Hor, Williams,
  Checkelsky, Roushan, Seo, Xu, Zandbergen, Yazdani, Ong, and
  Cava}}]{hor2010superconductivity}
\bibinfo{author}{\bibfnamefont{Y.~S.} \bibnamefont{Hor}},
  \bibinfo{author}{\bibfnamefont{A.~J.} \bibnamefont{Williams}},
  \bibinfo{author}{\bibfnamefont{J.~G.} \bibnamefont{Checkelsky}},
  \bibinfo{author}{\bibfnamefont{P.}~\bibnamefont{Roushan}},
  \bibinfo{author}{\bibfnamefont{J.}~\bibnamefont{Seo}},
  \bibinfo{author}{\bibfnamefont{Q.}~\bibnamefont{Xu}},
  \bibinfo{author}{\bibfnamefont{H.~W.} \bibnamefont{Zandbergen}},
  \bibinfo{author}{\bibfnamefont{A.}~\bibnamefont{Yazdani}},
  \bibinfo{author}{\bibfnamefont{N.~P.} \bibnamefont{Ong}}, \bibnamefont{and}
  \bibinfo{author}{\bibfnamefont{R.~J.} \bibnamefont{Cava}},
  \bibinfo{journal}{Physical review letters} \textbf{\bibinfo{volume}{104}},
  \bibinfo{pages}{057001} (\bibinfo{year}{2010}).

\bibitem[{\citenamefont{Wray et~al.}(2010)\citenamefont{Wray, Xu, Xia, San~Hor,
  Qian, Fedorov, Lin, Bansil, Cava, and Hasan}}]{wray2010observation}
\bibinfo{author}{\bibfnamefont{L.~A.} \bibnamefont{Wray}},
  \bibinfo{author}{\bibfnamefont{S.-Y.} \bibnamefont{Xu}},
  \bibinfo{author}{\bibfnamefont{Y.}~\bibnamefont{Xia}},
  \bibinfo{author}{\bibfnamefont{Y.}~\bibnamefont{San~Hor}},
  \bibinfo{author}{\bibfnamefont{D.}~\bibnamefont{Qian}},
  \bibinfo{author}{\bibfnamefont{A.~V.} \bibnamefont{Fedorov}},
  \bibinfo{author}{\bibfnamefont{H.}~\bibnamefont{Lin}},
  \bibinfo{author}{\bibfnamefont{A.}~\bibnamefont{Bansil}},
  \bibinfo{author}{\bibfnamefont{R.~J.} \bibnamefont{Cava}}, \bibnamefont{and}
  \bibinfo{author}{\bibfnamefont{M.~Z.} \bibnamefont{Hasan}},
  \bibinfo{journal}{Nature Physics} \textbf{\bibinfo{volume}{6}},
  \bibinfo{pages}{855} (\bibinfo{year}{2010}).

\bibitem[{\citenamefont{Kriener et~al.}(2011)\citenamefont{Kriener, Segawa,
  Ren, Sasaki, and Ando}}]{kriener2011bulk}
\bibinfo{author}{\bibfnamefont{M.}~\bibnamefont{Kriener}},
  \bibinfo{author}{\bibfnamefont{K.}~\bibnamefont{Segawa}},
  \bibinfo{author}{\bibfnamefont{Z.}~\bibnamefont{Ren}},
  \bibinfo{author}{\bibfnamefont{S.}~\bibnamefont{Sasaki}}, \bibnamefont{and}
  \bibinfo{author}{\bibfnamefont{Y.}~\bibnamefont{Ando}},
  \bibinfo{journal}{Physical Review Letters} \textbf{\bibinfo{volume}{106}},
  \bibinfo{pages}{127004} (\bibinfo{year}{2011}).

\bibitem[{\citenamefont{Shruti et~al.}(2015)\citenamefont{Shruti, Maurya, Neha,
  Srivastava, and Patnaik}}]{Shruti_SrBiSe}
\bibinfo{author}{\bibnamefont{Shruti}}, \bibinfo{author}{\bibfnamefont{V.~K.}
  \bibnamefont{Maurya}},
  \bibinfo{author}{\bibfnamefont{P.}~\bibnamefont{Neha}},
  \bibinfo{author}{\bibfnamefont{P.}~\bibnamefont{Srivastava}},
  \bibnamefont{and} \bibinfo{author}{\bibfnamefont{S.}~\bibnamefont{Patnaik}},
  \bibinfo{journal}{Phys. Rev. B} \textbf{\bibinfo{volume}{92}},
  \bibinfo{pages}{020506} (\bibinfo{year}{2015}),
  \urlprefix\url{https://link.aps.org/doi/10.1103/PhysRevB.92.020506}.

\bibitem[{\citenamefont{Liu et~al.}(2015)\citenamefont{Liu, Yao, Shao, Zuo, Pi,
  Tan, Zhang, and Zhang}}]{Liu_SrBiSe}
\bibinfo{author}{\bibfnamefont{Z.}~\bibnamefont{Liu}},
  \bibinfo{author}{\bibfnamefont{X.}~\bibnamefont{Yao}},
  \bibinfo{author}{\bibfnamefont{J.}~\bibnamefont{Shao}},
  \bibinfo{author}{\bibfnamefont{M.}~\bibnamefont{Zuo}},
  \bibinfo{author}{\bibfnamefont{L.}~\bibnamefont{Pi}},
  \bibinfo{author}{\bibfnamefont{S.}~\bibnamefont{Tan}},
  \bibinfo{author}{\bibfnamefont{C.}~\bibnamefont{Zhang}}, \bibnamefont{and}
  \bibinfo{author}{\bibfnamefont{Y.}~\bibnamefont{Zhang}},
  \bibinfo{journal}{Journal of the American Chemical Society}
  \textbf{\bibinfo{volume}{137}}, \bibinfo{pages}{10512}
  (\bibinfo{year}{2015}),
  \urlprefix\url{http://dx.doi.org/10.1021/jacs.5b06815}.

\bibitem[{\citenamefont{Pan et~al.}(2016)\citenamefont{Pan, Nikitin, Araizi,
  Huang, Matsushita, Naka, and de~Visser}}]{Pan_Hc2}
\bibinfo{author}{\bibfnamefont{Y.}~\bibnamefont{Pan}},
  \bibinfo{author}{\bibfnamefont{A.~M.} \bibnamefont{Nikitin}},
  \bibinfo{author}{\bibfnamefont{G.~K.} \bibnamefont{Araizi}},
  \bibinfo{author}{\bibfnamefont{Y.~K.} \bibnamefont{Huang}},
  \bibinfo{author}{\bibfnamefont{Y.}~\bibnamefont{Matsushita}},
  \bibinfo{author}{\bibfnamefont{T.}~\bibnamefont{Naka}}, \bibnamefont{and}
  \bibinfo{author}{\bibfnamefont{A.}~\bibnamefont{de~Visser}},
  \bibinfo{journal}{Scientific Reports} \textbf{\bibinfo{volume}{6}},
  \bibinfo{pages}{28632} (\bibinfo{year}{2016}),
  \urlprefix\url{http://dx.doi.org/10.1038/srep28632}.

\bibitem[{\citenamefont{Matano et~al.}(2016)\citenamefont{Matano, Kriener,
  Segawa, Ando, and Zheng}}]{Matano.Kriener.ea:16}
\bibinfo{author}{\bibfnamefont{K.}~\bibnamefont{Matano}},
  \bibinfo{author}{\bibfnamefont{M.}~\bibnamefont{Kriener}},
  \bibinfo{author}{\bibfnamefont{K.}~\bibnamefont{Segawa}},
  \bibinfo{author}{\bibfnamefont{Y.}~\bibnamefont{Ando}}, \bibnamefont{and}
  \bibinfo{author}{\bibfnamefont{G.-q.} \bibnamefont{Zheng}},
  \bibinfo{journal}{Nat Phys} \textbf{\bibinfo{volume}{12}},
  \bibinfo{pages}{852} (\bibinfo{year}{2016}), ISSN \bibinfo{issn}{1745-2473},
  \urlprefix\url{http://dx.doi.org/10.1038/nphys3781}.

\bibitem[{\citenamefont{Tao et~al.}(2018)\citenamefont{Tao, Yan, Liu, Wang,
  Ando, Wang, Zhang, and Feng}}]{tao2018direct}
\bibinfo{author}{\bibfnamefont{R.}~\bibnamefont{Tao}},
  \bibinfo{author}{\bibfnamefont{Y.-J.} \bibnamefont{Yan}},
  \bibinfo{author}{\bibfnamefont{X.}~\bibnamefont{Liu}},
  \bibinfo{author}{\bibfnamefont{Z.-W.} \bibnamefont{Wang}},
  \bibinfo{author}{\bibfnamefont{Y.}~\bibnamefont{Ando}},
  \bibinfo{author}{\bibfnamefont{Q.-H.} \bibnamefont{Wang}},
  \bibinfo{author}{\bibfnamefont{T.}~\bibnamefont{Zhang}}, \bibnamefont{and}
  \bibinfo{author}{\bibfnamefont{D.-L.} \bibnamefont{Feng}},
  \bibinfo{journal}{Physical Review X} \textbf{\bibinfo{volume}{8}},
  \bibinfo{pages}{041024} (\bibinfo{year}{2018}).

\bibitem[{\citenamefont{Asaba et~al.}(2017)\citenamefont{Asaba, Lawson,
  Tinsman, Chen, Corbae, Li, Qiu, Hor, Fu, and Li}}]{Asaba_NbBiSe}
\bibinfo{author}{\bibfnamefont{T.}~\bibnamefont{Asaba}},
  \bibinfo{author}{\bibfnamefont{B.~J.} \bibnamefont{Lawson}},
  \bibinfo{author}{\bibfnamefont{C.}~\bibnamefont{Tinsman}},
  \bibinfo{author}{\bibfnamefont{L.}~\bibnamefont{Chen}},
  \bibinfo{author}{\bibfnamefont{P.}~\bibnamefont{Corbae}},
  \bibinfo{author}{\bibfnamefont{G.}~\bibnamefont{Li}},
  \bibinfo{author}{\bibfnamefont{Y.}~\bibnamefont{Qiu}},
  \bibinfo{author}{\bibfnamefont{Y.~S.} \bibnamefont{Hor}},
  \bibinfo{author}{\bibfnamefont{L.}~\bibnamefont{Fu}}, \bibnamefont{and}
  \bibinfo{author}{\bibfnamefont{L.}~\bibnamefont{Li}}, \bibinfo{journal}{Phys.
  Rev. X} \textbf{\bibinfo{volume}{7}}, \bibinfo{pages}{011009}
  (\bibinfo{year}{2017}),
  \urlprefix\url{https://link.aps.org/doi/10.1103/PhysRevX.7.011009}.

\bibitem[{\citenamefont{Speight et~al.}(2022)\citenamefont{Speight, Winyard,
  and Babaev}}]{nematic}
\bibinfo{author}{\bibfnamefont{M.}~\bibnamefont{Speight}},
  \bibinfo{author}{\bibfnamefont{T.}~\bibnamefont{Winyard}}, \bibnamefont{and}
  \bibinfo{author}{\bibfnamefont{E.}~\bibnamefont{Babaev}},
  \emph{\bibinfo{title}{Symmetries, length scales, magnetic response and
  skyrmion chains in nematic superconductors}} (\bibinfo{year}{2022}),
  \eprint{arXiv:2202.13674}.

\bibitem[{\citenamefont{How and Yip}(2020{\natexlab{b}})}]{how2020half}
\bibinfo{author}{\bibfnamefont{P.~T.} \bibnamefont{How}} \bibnamefont{and}
  \bibinfo{author}{\bibfnamefont{S.-K.} \bibnamefont{Yip}},
  \bibinfo{journal}{Physical Review Research} \textbf{\bibinfo{volume}{2}},
  \bibinfo{pages}{043192} (\bibinfo{year}{2020}{\natexlab{b}}).

\bibitem[{\citenamefont{Silaev et~al.}(2018)\citenamefont{Silaev, Winyard, and
  Babaev}}]{silaev2018non}
\bibinfo{author}{\bibfnamefont{M.}~\bibnamefont{Silaev}},
  \bibinfo{author}{\bibfnamefont{T.}~\bibnamefont{Winyard}}, \bibnamefont{and}
  \bibinfo{author}{\bibfnamefont{E.}~\bibnamefont{Babaev}},
  \bibinfo{journal}{Physical Review B} \textbf{\bibinfo{volume}{97}},
  \bibinfo{pages}{174504} (\bibinfo{year}{2018}).

\bibitem[{\citenamefont{Winyard
  et~al.}(2019{\natexlab{a}})\citenamefont{Winyard, Silaev, and
  Babaev}}]{winyard2019hierarchies}
\bibinfo{author}{\bibfnamefont{T.}~\bibnamefont{Winyard}},
  \bibinfo{author}{\bibfnamefont{M.}~\bibnamefont{Silaev}}, \bibnamefont{and}
  \bibinfo{author}{\bibfnamefont{E.}~\bibnamefont{Babaev}},
  \bibinfo{journal}{Physical Review B} \textbf{\bibinfo{volume}{99}},
  \bibinfo{pages}{064509} (\bibinfo{year}{2019}{\natexlab{a}}).

\bibitem[{\citenamefont{Winyard
  et~al.}(2019{\natexlab{b}})\citenamefont{Winyard, Silaev, and
  Babaev}}]{winyard2019skyrmion}
\bibinfo{author}{\bibfnamefont{T.}~\bibnamefont{Winyard}},
  \bibinfo{author}{\bibfnamefont{M.}~\bibnamefont{Silaev}}, \bibnamefont{and}
  \bibinfo{author}{\bibfnamefont{E.}~\bibnamefont{Babaev}},
  \bibinfo{journal}{Physical Review B} \textbf{\bibinfo{volume}{99}},
  \bibinfo{pages}{024501} (\bibinfo{year}{2019}{\natexlab{b}}).

\bibitem[{\citenamefont{Speight et~al.}(2019)\citenamefont{Speight, Winyard,
  and Babaev}}]{speight2019chiral}
\bibinfo{author}{\bibfnamefont{M.}~\bibnamefont{Speight}},
  \bibinfo{author}{\bibfnamefont{T.}~\bibnamefont{Winyard}}, \bibnamefont{and}
  \bibinfo{author}{\bibfnamefont{E.}~\bibnamefont{Babaev}},
  \bibinfo{journal}{Physical Review B} \textbf{\bibinfo{volume}{100}},
  \bibinfo{pages}{174514} (\bibinfo{year}{2019}).

\bibitem[{\citenamefont{Sonier et~al.}(2000)\citenamefont{Sonier, Brewer, and
  Kiefl}}]{sonier2000musr}
\bibinfo{author}{\bibfnamefont{J.~E.} \bibnamefont{Sonier}},
  \bibinfo{author}{\bibfnamefont{J.~H.} \bibnamefont{Brewer}},
  \bibnamefont{and} \bibinfo{author}{\bibfnamefont{R.~F.} \bibnamefont{Kiefl}},
  \bibinfo{journal}{Reviews of Modern Physics} \textbf{\bibinfo{volume}{72}},
  \bibinfo{pages}{769} (\bibinfo{year}{2000}).

\bibitem[{\citenamefont{Venderbos
  et~al.}(2016{\natexlab{b}})\citenamefont{Venderbos, Kozii, and
  Fu}}]{venderbos2016identification}
\bibinfo{author}{\bibfnamefont{J.~W.} \bibnamefont{Venderbos}},
  \bibinfo{author}{\bibfnamefont{V.}~\bibnamefont{Kozii}}, \bibnamefont{and}
  \bibinfo{author}{\bibfnamefont{L.}~\bibnamefont{Fu}},
  \bibinfo{journal}{Physical Review B} \textbf{\bibinfo{volume}{94}},
  \bibinfo{pages}{094522} (\bibinfo{year}{2016}{\natexlab{b}}).

\bibitem[{\citenamefont{Parzen}(1962)}]{parzen1962estimation}
\bibinfo{author}{\bibfnamefont{E.}~\bibnamefont{Parzen}}, \bibinfo{journal}{The
  annals of mathematical statistics} \textbf{\bibinfo{volume}{33}},
  \bibinfo{pages}{1065} (\bibinfo{year}{1962}).

\bibitem[{\citenamefont{Biswas et~al.}(2020)\citenamefont{Biswas, Rybakov,
  Singh, Mukherjee, Parzyk, Balakrishnan, Lees, Dewhurst, Babaev, Hillier
  et~al.}}]{biswas2020coexistence}
\bibinfo{author}{\bibfnamefont{P.}~\bibnamefont{Biswas}},
  \bibinfo{author}{\bibfnamefont{F.~N.} \bibnamefont{Rybakov}},
  \bibinfo{author}{\bibfnamefont{R.}~\bibnamefont{Singh}},
  \bibinfo{author}{\bibfnamefont{S.}~\bibnamefont{Mukherjee}},
  \bibinfo{author}{\bibfnamefont{N.}~\bibnamefont{Parzyk}},
  \bibinfo{author}{\bibfnamefont{G.}~\bibnamefont{Balakrishnan}},
  \bibinfo{author}{\bibfnamefont{M.}~\bibnamefont{Lees}},
  \bibinfo{author}{\bibfnamefont{C.}~\bibnamefont{Dewhurst}},
  \bibinfo{author}{\bibfnamefont{E.}~\bibnamefont{Babaev}},
  \bibinfo{author}{\bibfnamefont{A.}~\bibnamefont{Hillier}},
  \bibnamefont{et~al.}, \bibinfo{journal}{Physical Review B}
  \textbf{\bibinfo{volume}{102}}, \bibinfo{pages}{144523}
  (\bibinfo{year}{2020}).

\bibitem[{\citenamefont{Ray et~al.}(2014)\citenamefont{Ray, Gibbs, Bending,
  Curran, Babaev, Baines, Mackenzie, and Lee}}]{ray2014muon}
\bibinfo{author}{\bibfnamefont{S.}~\bibnamefont{Ray}},
  \bibinfo{author}{\bibfnamefont{A.}~\bibnamefont{Gibbs}},
  \bibinfo{author}{\bibfnamefont{S.}~\bibnamefont{Bending}},
  \bibinfo{author}{\bibfnamefont{P.}~\bibnamefont{Curran}},
  \bibinfo{author}{\bibfnamefont{E.}~\bibnamefont{Babaev}},
  \bibinfo{author}{\bibfnamefont{C.}~\bibnamefont{Baines}},
  \bibinfo{author}{\bibfnamefont{A.}~\bibnamefont{Mackenzie}},
  \bibnamefont{and} \bibinfo{author}{\bibfnamefont{S.}~\bibnamefont{Lee}},
  \bibinfo{journal}{Physical Review B} \textbf{\bibinfo{volume}{89}},
  \bibinfo{pages}{094504} (\bibinfo{year}{2014}).

\bibitem[{\citenamefont{Speight et~al.}(2021)\citenamefont{Speight, Winyard,
  Wormald, and Babaev}}]{speight2021magnetic}
\bibinfo{author}{\bibfnamefont{M.}~\bibnamefont{Speight}},
  \bibinfo{author}{\bibfnamefont{T.}~\bibnamefont{Winyard}},
  \bibinfo{author}{\bibfnamefont{A.}~\bibnamefont{Wormald}}, \bibnamefont{and}
  \bibinfo{author}{\bibfnamefont{E.}~\bibnamefont{Babaev}},
  \bibinfo{journal}{Physical Review B} \textbf{\bibinfo{volume}{104}},
  \bibinfo{pages}{174515} (\bibinfo{year}{2021}).

\bibitem[{\citenamefont{Benfenati et~al.}(2020)\citenamefont{Benfenati,
  Barkman, Winyard, Wormald, Speight, and Babaev}}]{benfenati2020magnetic}
\bibinfo{author}{\bibfnamefont{A.}~\bibnamefont{Benfenati}},
  \bibinfo{author}{\bibfnamefont{M.}~\bibnamefont{Barkman}},
  \bibinfo{author}{\bibfnamefont{T.}~\bibnamefont{Winyard}},
  \bibinfo{author}{\bibfnamefont{A.}~\bibnamefont{Wormald}},
  \bibinfo{author}{\bibfnamefont{M.}~\bibnamefont{Speight}}, \bibnamefont{and}
  \bibinfo{author}{\bibfnamefont{E.}~\bibnamefont{Babaev}},
  \bibinfo{journal}{Physical Review B} \textbf{\bibinfo{volume}{101}},
  \bibinfo{pages}{054507} (\bibinfo{year}{2020}).

\bibitem[{\citenamefont{Speight and Winyard}(2020)}]{speight2020skyrmions}
\bibinfo{author}{\bibfnamefont{M.}~\bibnamefont{Speight}} \bibnamefont{and}
  \bibinfo{author}{\bibfnamefont{T.}~\bibnamefont{Winyard}},
  \bibinfo{journal}{Physical Review B} \textbf{\bibinfo{volume}{101}},
  \bibinfo{pages}{134420} (\bibinfo{year}{2020}).

\bibitem[{\citenamefont{Tinkham}(2004)}]{tinkham2004introduction}
\bibinfo{author}{\bibfnamefont{M.}~\bibnamefont{Tinkham}},
  \emph{\bibinfo{title}{Introduction to superconductivity}}
  (\bibinfo{publisher}{Courier Corporation}, \bibinfo{year}{2004}).

\end{thebibliography}

\end{document}